\documentclass[12pt,twoside]{article}

\usepackage{pslatex}    
\usepackage{amsfonts}
\usepackage{amsmath}
\usepackage{amsthm}
\usepackage{amscd}
\usepackage{cite}
\usepackage{eucal}
\usepackage{epsf}
\usepackage{epsfig}
\usepackage{a4}
\usepackage{booktabs}

\def\beq{\begin{equation}}
\def\eeq{\end{equation}}
\def\bea{\begin{eqnarray}}
\def\eea{\end{eqnarray}}

\def\beann{\begin{eqnarray*}}
\def\eeann{\end{eqnarray*}}

\let\a=\alpha \let\be=\beta \let\g=\gamma \let\de=\delta
\let\e=\varepsilon \let\z=\zeta \let\h=\eta 
\let\eps=\epsilon
  \let\la=\lambda \let\m=\mu
 \let\x=\xi \let\p=\pi \let\r=\rho \let\s=\sigma
\let\om=\omega \let\ps=\psi
\let\ph=\varphi  \let\PH=\Phi \let\Ps=\Psi
\let\Om=\Omega \let\Si=\Sigma \let\Th=\Theta
\let\La=\Lambda \let\G=\Gamma \let\D=\Delta

\newcommand{\bare}{\bar \varepsilon}

\let\qd=\quad \let\qqd=\qquad 

\def\epp{\, .}
\def\epc{\, ,}

\def\tst#1{{\textstyle #1}}

\theoremstyle{plain}

\newtheorem*{corollary*}{Corollary}

\newtheorem*{conjecture*}{Conjecture}

\theoremstyle{definition}

\def\2{\frac{1}{2}} \def\4{\frac{1}{4}}

\def\6{\partial}

\def\+{\dagger}

\def\<{\langle} \def\>{\rangle}

\let\auf=\uparrow \let\ab=\downarrow

\def\i{{\rm i}}

\def\rd{{\rm d}}
\def\re{{\rm e}}

\DeclareMathOperator{\sh}{sh}
\DeclareMathOperator{\ch}{ch}
\DeclareMathOperator{\tgh}{th}

\DeclareMathOperator{\tr}{tr}

\DeclareMathOperator{\End}{End}

\def\Re{{\rm Re\,}} \def\Im{{\rm Im\,}}

\def\diag{{\rm diag}}

\def\vs{\mathbf{s}}

\def\fa{\mathfrak{a}}

\def\faq{\overline{\mathfrak{a}}}
\def\fbq{\overline{\mathfrak{b}}}

\def\fb{\mathfrak{b}}

\renewcommand{\appendix}{%
   \renewcommand{\section}{
        \secdef\Appendix\sAppendix}%
   \setcounter{section}{0}%
   \renewcommand{\thesection}{\Alph{section}}%
   \renewcommand{\theequation}{\thesection.\arabic{equation}}%
}

\newcommand{\Appendix}[2][?]{%
     \refstepcounter{section}%
     \setcounter{equation}{0}%
     \addcontentsline{toc}{appendix}%
          {\protect\numberline{\appendixname~\thesection} #1}%
     \vspace{\baselineskip}%
     {\noindent\large\bfseries\appendixname: #2\par}%
     \sectionmark{#1}\vspace{\baselineskip}}

\newcommand{\sAppendix}[1]{%
     {\noindent\large\bfseries\appendixname\:: #1\par}%
     \sectionmark{#1}\vspace{\baselineskip}}

\begin{document}

\thispagestyle{empty}

\begin{center}

{\Large {\bf Density matrices for finite segments of Heisenberg
chains of arbitrary length\\}}

\vspace{7mm}

{\large
Jens Damerau\footnote{e-mail: damerau@physik.uni-wuppertal.de},
Frank G\"{o}hmann\footnote{e-mail: goehmann@physik.uni-wuppertal.de},
Nils P. Hasenclever\footnote{e-mail:
hasenclever@physik.uni-wuppertal.de},\\
Andreas Kl\"umper\footnote{e-mail: kluemper@physik.uni-wuppertal.de}\\

\vspace{5mm}

Fachbereich C -- Physik, Bergische Universit\"at Wuppertal,\\
42097 Wuppertal, Germany\\}

\vspace{20mm}

{\large {\bf Abstract}}

\end{center}

\begin{list}{}{\addtolength{\rightmargin}{9mm}
               \addtolength{\topsep}{-5mm}}
\item
We derive a multiple integral representing the ground state density
matrix of a segment of length $m$ of the XXZ spin chain on $L$
lattice sites, which depends on $L$ only parametrically. This allows
us to treat chains of arbitrary finite length. Specializing to the
isotropic limit of the XXX chain we show for small $m$ that the
multiple integrals factorize. We conjecture that this property holds
for arbitrary $m$ and suggest an exponential formula for the density
matrix which involves only a double Cauchy type integral in the
exponent. We demonstrate the efficiency of our formula by computing
the next-to-nearest neighbour $zz$-correlation function for chain
lengths ranging from two to macroscopic numbers.\\[2ex]
\end{list}

\clearpage

\section{Introduction}
Formally, integrable systems at finite temperature have much in common
with finite-length systems. In the former case the free energy
in the thermodynamic limit can be expressed as the logarithm
of the dominant eigenvalue of a quantum transfer matrix
\cite{Suzuki85,SuIn87}, whereas in the latter case the logarithmic
derivative of the largest eigenvalue of the usual row-to-row transfer
matrix \cite{Babook} determines the ground state energy of the
system of length $L$. In both cases the technique of non-linear
integral equations \cite{KluBat90,KluBatPea91} can be applied to
express the transfer matrix eigenvalue as an integral over appropriately
defined auxiliary functions. The integrals can be evaluated numerically
yielding high precision data for thermodynamic properties at arbitrary
temperatures, or for the ground state energy at arbitrary lengths,
respectively.

Here we show for the XXZ spin-1/2 chain that this formal similarity
persists for a multiple integral representation of the density matrix
of a chain segment which was first derived for the ground state of
the infinitely long chain \cite{JMMN92,JiMi96,KMT99b} and later
generalized to finite temperature \cite{GKS05,GHS05}. We also show
examples which suggest that the factorization of the multiple integrals,
that was proven for the ground state of the infinite chain at
vanishing magnetic field \cite{BoKo01,BJMST04a,BJMST04b} and recently
observed at finite temperature and non-zero magnetic field
\cite{BGKS06}, might also generally hold for the ground state of a
finite chain.
\section{The XXZ chain and its integrable structure}
The XXZ chain is an anisotropic generalization of the Heisenberg spin
chain. If the value of all local spins is 1/2 the model is integrable,
and its Hamiltonian can be expressed through the local action of
the Pauli matrices $\s^x, \s^y, \s^z$ on $L$ sites of a chain,
\begin{equation} \label{xxzham}
     H = J \sum_{j=1}^L \Bigl( \s_{j-1}^x \s_j^x
           + \s_{j-1}^y \s_j^y + \D (\s_{j-1}^z \s_j^z - 1) \Bigr) \epp
\end{equation}
This Hamiltonian depends on two real parameters, the exchange coupling
$J$ and an aniso\-tro\-py parameter $\D$. We shall consider the critical
antiferromagnetic regime $J > 0$, $- 1 < \D \le 1$. Setting $\Th =
\diag(\re^{\i \Phi}, \re^{- \i \Phi})$, $\Phi \in [0, 2\p]$, we fix
the boundary conditions requiring that
\begin{equation} \label{twistbound}
     \begin{pmatrix} {e_0}_1^1 & {e_0}_2^1 \\
                     {e_0}_1^2 & {e_0}_2^2 \end{pmatrix} =
     \Th \begin{pmatrix} {e_L}_1^1 & {e_L}_2^1 \\
                         {e_L}_1^2 & {e_L}_2^2 \end{pmatrix} \Th^{-1}
			 \epc
\end{equation}
where the $e^\a_\be$, $\a, \be = 1, 2$, denote the gl(2) standard
basis ($\s^x = e_2^1 + e_1^2$, $\s^y = \i (e_2^1 - e_1^2)$, $\s^z =
e_1^1 - e_2^2$). We call $\Phi$ the twist angle. $\Phi = 0$ corresponds
to the familiar periodic boundary conditions.

All results in this paper rely heavily on the fact that $H$
can be derived from the well-known trigonometric $R$-matrix
\begin{align} \label{rxxz}
     R(\la) & = \begin{pmatrix}
                    1 & 0 & 0 & 0 \\
		    0 & b(\la) & c(\la) & 0 \\
		    0 & c(\la) & b(\la) & 0 \\
		    0 & 0 & 0 & 1
		\end{pmatrix} \epc \\[2ex]
     b(\la) & = \frac{\sh(\la)}{\sh(\la + \h)} \epc \qd
     c(\la) = \frac{\sh(\h)}{\sh(\la + \h)} \epc \label{defbc}
\end{align}
of the six-vertex model \cite{Babook}. Associating a $2 \times 2$
L-matrix with elements
\begin{equation} \label{deflmatrix}
     {L_j}^\a_\be (\la) =
          R^{\a \g}_{\be \de} (\la) {e_j}_\g^\de
\end{equation}
with every lattice site we can define the monodromy matrix of the
XXZ chain,
\begin{equation} \label{defmono}
     T(\la) = \begin{pmatrix}
                 A(\la) & B(\la) \\ C(\la) & D(\la)
              \end{pmatrix} =
              \Th L_L (\la) \dots L_1 (\la) \epp
\end{equation}
It provides, by construction, a representation of the Yang-Baxter
algebra,
\begin{equation} \label{rtt}
     \check R(\la - \m) \big( T (\la) \otimes T (\m) \big) =
        \big( T (\m) \otimes T (\la) \big) \check R(\la - \m) \epc
\end{equation}
where $\check R = P R$ if $P$ is the transposition of the two
factors in ${\mathbb C}^2 \otimes {\mathbb C}^2$. We define
the twisted transfer matrix $t(\la) = \tr T(\la)$. Then, due to
(\ref{rtt}), the function $\ln \bigl( t^{-1}(0) t(\la) \bigr)$
generates a sequence of commuting local operators. The first one
is proportional to the Hamiltonian (\ref{xxzham}),
\begin{equation} \label{hamfromt}
     H = 2 J \sh(\h) \,
         \6_\la \ln \bigl( t^{-1}(0) t(\la) \bigr)\big|_{\la = 0} \epc
\end{equation}
with twisted boundary conditions (\ref{twistbound}) if we identify
$\D = \ch (\h)$. The critical regime corresponds to purely imaginary
$\h = \i \g$, $\g \in [0, \p)$. Because of (\ref{hamfromt}) we may
solve the eigenvalue problem of the transfer matrix rather than
dealing directly with the Hamiltonian.
\section{The Bethe ansatz solution for the ground state}
\label{sec:bas}
The twisted transfer matrix $t(\la)$ can be diagonalized by means
of the algebraic Bethe ansatz. Since this technique has been explained
elsewhere (see e.g.\ \cite{KBIBo}), we may be content here with
a mere description of the result. Eigenstates $|\{\la\}\>$ of $t(\la)$
are generated by the multiple action of the operators $B(\la)$,
equation~(\ref{defmono}), on the ferromagnetic state $|0\> = \bigl(
\begin{smallmatrix} 1 \\ 0 \end{smallmatrix} \bigr)^{\otimes L}$,
\begin{equation} \label{bas}
     |\{\la\}\> = B \bigl(\la_1 -\tst{\frac{\h}{2}} \bigr)
                  \dots B \bigl(\la_N -\tst{\frac{\h}{2}} \bigr) |0\>
		  \epp
\end{equation}
Here the set of so-called Bethe roots $\{\la\} = \{\la_j\}_{j=1}^N$
is not arbitrary, but must be determined from the Bethe ansatz
equations
\begin{equation} \label{bae}
     1 + \frac{\re^{- 2\i \Phi} \sh^L (\la_j - \frac{\h}{2})}
              {\sh^L (\la_j + \frac{\h}{2})}
     \prod_{k=1}^N \frac{\sh(\la_j - \la_k + \h)}
                        {\sh(\la_j - \la_k - \h)} = 0 \epc
     \qd j = 1, \dots, N \epp
\end{equation}
The transfer matrix eigenvalues corresponding to the eigenstates
(\ref{bas}) are given by
\begin{equation} \label{eigenvalue}
     \La (\la) = \re^{\i \Phi} \prod_{j=1}^N
                 \frac{\sh \bigl(\la - \la_j - \frac{\h}{2} \bigr)}
                      {\sh \bigl(\la - \la_j + \frac{\h}{2} \bigr)}
               + \frac{\re^{- \i \Phi} \sh^L (\la)}{\sh^L (\la + \h)}
	         \prod_{j=1}^N
                 \frac{\sh \bigl(\la - \la_j + \frac{3\h}{2} \bigr)}
                      {\sh \bigl(\la - \la_j + \frac{\h}{2} \bigr)} \epp
\end{equation}

\begin{figure}
    \centering
    \includegraphics{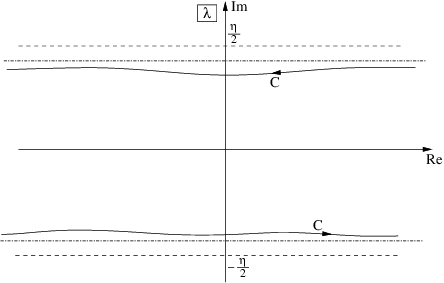}
    \caption{\label{fig:cont} The canonical contour ${\cal C}$ surrounds
             the real axis in counterclockwise manner inside the
             strip $- \frac{|\h|}{2} < \Im \la < \frac{|\h|}{2}$.} 
\end{figure}

In the following we shall concentrate on the ground state properties
of the XXZ chain. This brings about severe simplifications. We
will be dealing with a single rather special solution of the Bethe
ansatz equations, and we do not have to touch the delicate
question under which circumstances the set of states (\ref{bas}) is
complete. Let us fix an even length $L$ of the chain. Then the ground
state of the Hamiltonian (\ref{xxzham}) is the transfer matrix
eigenstate (\ref{bas}) with $\{\la_j\}_{j=1}^{L/2}$ the unique
real solution of (\ref{bae}) for $N = L/2$. The real and mutually
distinct Bethe roots in this solution uniquely determine a meromorphic
auxiliary function
\begin{equation} \label{auxa}
     \fa (\la) =
        \frac{\re^{- 2\i \Phi} \sh^L (\la - \frac{\h}{2})}
             {\sh^L (\la + \frac{\h}{2})}
     \prod_{k=1}^{L/2} \frac{\sh(\la - \la_k + \h)}
                            {\sh(\la - \la_k - \h)}
\end{equation}
in the complex plane. In terms of this function the ground state
eigenvalue $\La_0 (\la)$ becomes
\begin{equation}
     \La_0 (\la) = \bigl( 1 + \fa (\la + \tst{\frac{\h}{2}}) \bigr) \,
                   \re^{\i \Phi} \prod_{j=1}^{L/2}
                   \frac{\sh \bigl(\la - \la_j - \frac{\h}{2} \bigr)}
                        {\sh \bigl(\la - \la_j + \frac{\h}{2} \bigr)}
			\epp
\end{equation}
It follows from the Bethe ansatz equations that $\La_0 (\la)$ is
regular at the points $\la_j - \frac{\h}{2}$, $j = 1, \dots, L/2$.
Extensive numerical studies moreover support the conjecture that
$\La_0 (\la)$ is non-zero inside a strip $- |\h| \le \Im \la \le 0$.
Our following treatment of the ground state of the finite-size
system is based on this conjecture. It has strong immediate
consequences. It implies that the function $1 + \fa(\la)$ is
analytic inside the strip $- \frac{|\h|}{2} < \Im \la \le
\frac{|\h|}{2}$ and that its only zeros in this strip are the
Bethe roots. This together with the obvious analytic
and asymptotic properties of $\fa (\la)$ is enough to set up a set
of functional equations for the second logarithmic derivatives
of $\fa (\la)$ and $1 + \fa(\la)$ which together with their
known asymptotics determine $\fa (\la)$ uniquely \cite{KluBatPea91}.
Once the functional equations are formulated it is easy to transform
them into non-linear integral equations. This procedure is, however,
non-unique. We usually work with two alternative forms of the
non-linear integral equations. One (we call it the `$\fa$-form') is
convenient for theoretical purposes. We shall use it in order to
write the density matrix as a multiple integral. The other form
(we call it the `$\fb \fbq$-form') is useful for numerical
calculations. The $\fb \fbq$-form was first derived in
\cite{KluBatPea91}. We show it later when we demonstrate the
numerical efficiency of our formulae. Here is the $\fa$-form of the
non-linear integral equation,
\begin{equation} \label{nlie}
     \ln \fa (\la) = - 2\i \Phi + L \h + L \ln \biggl(
                       \frac{\sh(\la - \frac{\h}{2})}
		            {\sh(\la + \frac{\h}{2})} \biggr)
                     - \int_{\cal C} \frac{\rd \om}{2 \p}
		       K_\h (\la - \om) \ln( 1 + \fa (\om)) \epp
\end{equation}
The integration contour ${\cal C}$ is shown in figure~\ref{fig:cont}.
It surrounds the real axis in counterclockwise manner. The
kernel $K_\h (\la)$ is defined as
\begin{equation}
     K_\h (\la) = \frac{\sh(2\h)}{\i \sh(\la - \h) \sh(\la + \h)} \epp
\end{equation}
We claim that inside the strip $- \frac{|\h|}{2} \le \Im \la \le
\frac{|\h|}{2}$ the function $\fa (\la)$ as defined in (\ref{nlie})
is the same as the function $\fa (\la)$ as defined in (\ref{auxa})
and (\ref{bae}) with $N = L/2$. The ground state eigenvalue
$\La_0 (\la)$ can also be expressed as an integral over $\fa (\la)$,
\begin{equation}
     \ln \La_0 (\la) = \i \Phi + \frac{L(\i \p - \h)}{2}
                       + \int_{\cal C} \frac{\rd \om}{2 \p}
		       K_\frac{\h}{2} (\la - \om) \ln( 1 + \fa (\om))
		       \epp
\end{equation}
This determines the ground state energy and the eigenvalues of
the higher conserved quantities as a function of $L$.

Another implication of the fact that the function $1 + \fa(\la)$
is analytic inside the strip $- \frac{|\h|}{2} < \Im \la \le
\frac{|\h|}{2}$ and that its only zeros in this strip are the Bethe
roots is the formula
\begin{equation} \label{sumtoint}
     \int_{\cal C} \frac{\rd \om}{2 \p \i} \frac{f(\om)}{1 + \fa (\om)}
        = \sum_{j=1}^{L/2} \frac{f(\la_j)}{\fa'(\la_j)}
\end{equation}
which holds for any function $f(\la)$ analytic in the strip. This
formula, when read from right to left, enables us to rewrite sums
over the ground state Bethe roots as integrals over the `canonical
contour' ${\cal C}$. It is one of the key tools in the derivation
of the integral representation for the density matrix shown in the
next section.
\section{The density matrix as a multiple integral}
The density matrix is a means to describe a sub-system as part of a
larger system in thermodynamic equilibrium in terms of the degrees of
freedom of the sub-system. Typically the system is divided into two
parts, interpreted as sub-system and environment. Then one is usually
interested in the limit when the sub-system is kept fixed and the
size of the environment goes to infinity. Here we shall keep both
parts finite and study the influence of the size of the environment
on the sub-system which will be a segment consisting of the first
$m < L$ lattice sites of the XXZ chain. The environment will consist
of the remaining sites.

Let
\begin{equation} \label{statoph}
      \r_L = \frac{\re^{- \frac{H}{T}}}
                  {\tr \re^{- \frac{H}{T}}}
\end{equation}
be the statistical operator for the chain at temperature $T$.
Then the density matrix of the sub-system consisting of the first
$m$ lattice sites is defined as
\begin{equation} \label{defdensmatgen}
     D_L (T) = \tr_{m+1 \dots L} \r_L \epp
\end{equation}
By construction, the thermal average of every operator $A$ acting
non-trivially only on sites 1 to $m$ can now be written as
\begin{equation}
     \<A\>_{T} = \tr_{1 \dots L} A \r_L
               = \tr_{1 \dots m} A_{1 \dots m}
	               \tr_{m+1 \dots L} \r_L
               = \tr_{1 \dots m} A_{1 \dots m} D_L (T) \epc
\end{equation}
where $A_{1 \dots m}$ is the restriction of $A$ to a chain consisting
of sites 1 to $m$. In particular, every two-point function of local
operators in the segment 1 to $m$ of the XXZ chain can be brought into
the above form. If we follow the common convention and use the same
symbols for the local operators ${e_j}^\a_\be$ and for their
restriction to the first $m$ sites, we find the expression
\begin{equation} \label{densmatell}
     {D_L}^{\a_1 \dots \a_m}_{\be_1 \dots \be_m} (T)
        = \tr_{1 \dots m} {e_1}^{\a_1}_{\be_1} \dots
                          {e_m}^{\a_m}_{\be_m} D_L (T)
        = \<{e_1}^{\a_1}_{\be_1} \dots {e_m}^{\a_m}_{\be_m}\>_{T}
\end{equation}
for the matrix elements of the density matrix.

Here we are interested in the unique (normalized) ground state 
$|\Ps_0\>$ of the system of finite even length. In the limit
$T \rightarrow 0+$ the statistical operator $\rho_L$ converges to
the projector $|\Ps_0\>\<\Ps_0|$ onto the ground state, and the
formula (\ref{densmatell}) for the density matrix elements turns
into
\begin{equation} \label{densmatelgs}
     {D_L}^{\a_1 \dots \a_m}_{\be_1 \dots \be_m} =
        \lim_{T \rightarrow 0+}
	{D_L}^{\a_1 \dots \a_m}_{\be_1 \dots \be_m} (T) =
        \<\Ps_0| {e_1}^{\a_1}_{\be_1} \dots {e_m}^{\a_m}_{\be_m}|\Ps_0\>
	\epp
\end{equation}

We shall use a trick suggested in \cite{KMT99a} in order to express
(\ref{densmatelgs}) entirely in terms of data related to the monodromy
matrix $T(\la)$. Setting $\la = 0$ in the second equation~%
(\ref{defmono}), using $L_j (0) = R_{0j} (0) = P_{0j}$ (if $0$ denotes
the auxiliary space) and comparing both sides \cite{GoKo00} we
obtain
\begin{equation} \label{invformula}
     {e_j}_\be^\a = t^{j - 1} (0) T_\be^\a (0) t^{-j} (0) \epp
\end{equation}
It follows that
\begin{equation}
     {D_L}^{\a_1 \dots \a_m}_{\be_1 \dots \be_m} =
        \<\Ps_0| T^{\a_1}_{\be_1} (0) \dots T^{\a_m}_{\be_m} (0)
	         t^{-m} (0) |\Ps_0\> \epp
\end{equation}
In order to apply the techniques developed in \cite{GKS04a} for the
finite-temperature case we regularize the expression by introducing
inhomogeneity parameters $\x_j$, $j = 1, \dots, m$, in the following
way. Define an `inhomogeneous density matrix'
\begin{equation} \label{definhomdensmat}
     {D_L}^{\a_1 \dots \a_m}_{\be_1 \dots \be_m} (\x_1, \dots, \x_m) =
        \frac{\<\{\la\}| T_{\be_1}^{\a_1} (\x_1 - \frac{\h}{2}) \dots
                         T_{\be_m}^{\a_m} (\x_m - \frac{\h}{2})
              |\{\la\}\>}{\<\{\la\}|\{\la\}\>
	                  \prod_{j=1}^m \La_0 (\x_j -\frac{\h}{2})} \epc
\end{equation}
where $|\{\la\}\>$ is the (unnormalized) Bethe ansatz ground state.
Then
\begin{equation}
     {D_L}^{\a_1 \dots \a_m}_{\be_1 \dots \be_m} =
        \lim_{\x_1, \dots, \x_m \rightarrow \frac{\h}{2}}
	{D_L}^{\a_1 \dots \a_m}_{\be_1 \dots \be_m} (\x_1, \dots, \x_m)
	\epp
\end{equation}

The expression (\ref{definhomdensmat}) is of the same form as in the
finite-temperature case considered in \cite{GKS05,GHS05} with the
monodromy matrix elements obeying the same commutation relations
(\ref{rtt}). Moreover we have an auxiliary function $\fa (\la)$
which shares some of the features of the finite-temperature auxiliary
function and satisfies, in particular, equation~(\ref{sumtoint}).
The inhomogeneous density matrix can therefore be represented as a
multiple integral following the same lines of reasoning as in
\cite{GHS05}. Since the calculations are very similar we can skip
all details here and present the final result.

Let $|\a^+|$ be the number of up-spins (or ones) in the sequence of
upper indices $(\a_j)_{j=1}^m$ of the inhomogeneous density matrix
element (\ref{definhomdensmat}) and $|\be^-|$ the number of down-spins
(or twos) in the sequence of lower indices. The conservation of the
$z$-component of the total spin implies that all density matrix
elements with $|\a^+| + |\be^-| \ne m$ must vanish. Hence, $|\be^-|
= m - |\a^+|$ for the non-vanishing density matrix elements. Those are
conveniently labeled by two finite sequences of positive integers
$(x_j)_{j=1}^{|\a^+|}$ and $(y_k)_{k = |\a^+| + 1}^{m}$, where $x_j$
denotes the position of the $(|\a^+| - j + 1)$th up-spin in
$(\a_j)_{j=1}^m$, and $y_k$ denotes the position of the $(k - |\a^+|)$th
down-spin in $(\be_j)_{j=1}^m$. Then
\begin{align} \label{densint}
     {D_L}^{\a_1 \dots \a_m}_{\be_1 \dots \be_m}(\x_1, \dots, & \x_m)
        = \notag \\
	  \biggl[ \prod_{j=1}^{|\a^+|}
	     \int_{\cal C} & \frac{d \om_j}{2 \p \i (1 + \fa (\om_j))}
	     \prod_{k=1}^{x_j - 1} \sh(\om_j - \x_k - \h)
	     \prod_{k = x_j + 1}^m \sh(\om_j - \x_k) \biggr]
	     \notag \\
          \biggl[ \prod_{j = |\a^+| + 1}^{m}
	     \int_{\cal C} & \frac{d \om_j}{2 \p \i (1 + \faq (\om_j))}
	     \prod_{k=1}^{y_j - 1} \sh(\om_j - \x_k + \h)
	     \prod_{k = y_j + 1}^m \sh(\om_j - \x_k) \biggr]
	     \notag \\ &
        \frac{\det( - G(\om_j, \x_k))}
	     {\prod_{1 \le j < k \le m}
	         \sh(\x_k - \x_j) \sh( \om_j - \om_k - \h)} \epc
\end{align}
where $\faq = 1/\fa$ and where the function $G(\om,\x)$ has to be
calculated from the linear integral equation
\begin{equation} \label{defg}
     G(\la,\x) = \frac{\sh(\h)}{\sh(\la - \x) \sh(\la - \x - \h)}
                 + \int_{\cal C}
		   \frac{d \om \: G (\om,\x)}{2 \p (1 + \fa (\om))} \,
                   K_\h (\la - \om) \epp
\end{equation}
The contour ${\cal C}$ in (\ref{densint}), (\ref{defg}) is the same as
in figure~\ref{fig:cont}. Remarkably, (\ref{densint}) and (\ref{defg})
are of precisely the same form as in the finite-temperature case, the
only difference being the definition of the auxiliary function $\fa$,
equation~(\ref{nlie}). Thus, many results that were obtained for the
finite-temperature case can be carried over to the finite-size case
without further effort.

Performing the homogeneous limit $\x_j \rightarrow \frac{\h}{2}$ in
(\ref{densint}) we obtain a multiple integral formula for the
density matrix ${D_L}^{\a_1 \dots \a_m}_{\be_1 \dots \be_m}$. This
limit was described elsewhere \cite{KMT99b,GKS05}. Here we have to
take into account that in the derivation of (\ref{densint}), (\ref{defg})
we have assumed that the $\x_j$ lie inside $\cal C$. Thus, in order to
calculate the homogeneous limit, we first have to push the contour
to $\pm \frac{\h}{2}$. It turns out that the multiple integral formula
for the homogeneous density matrix is not very efficient numerically
(see \cite{BoGo05} for the finite-temperature case). For this reason
and for space limitations we leave the homogeneous limit of the multiple
integral as an exercise to the reader. In the next section we shall
rather proceed along the lines of the recent paper \cite{BGKS06} where
for the isotropic model the inhomogeneous formula (\ref{densint}) was
first split into a sum over products of single integrals and where the
homogeneous limit was performed only after that. To be more precise,
such `factorization' was carried out for $m = 2, 3$ and then a general
formula inspired by \cite{BJMST05b} was conjectured for finite
temperature but zero magnetic field.
\section{Factorization for XXX}
In the following we restrict ourselves to the isotropic limit
$\D \rightarrow 1$. In order to perform this limit in our formulae
we have to replace $\h$ by $\i \eps$ with $\eps \rightarrow 0$. In
a similar way we have to rescale the spectral parameter $\la$
in (\ref{nlie}), (\ref{defg}), the inhomogeneities, the integration
variables and the functions $\fa$ and $G$. Then
\begin{align} \label{densintxxx}
     {D_L}^{\a_1 \dots \a_m}_{\be_1 \dots \be_m} (\x_1, \dots, \x_m)
        = & \biggl[ \prod_{j=1}^{|\a^+|}
             \int_{\cal C} \frac{\rd \om_j}{2 \p (1 + \fa (\om_j))}
             \prod_{k=1}^{x_j - 1} (\om_j - \x_k - \i)
             \prod_{k = x_j + 1}^m (\om_j - \x_k) \biggr] \notag \\
          & \biggl[ \prod_{j = |\a^+| + 1}^{m}
             \int_{\cal C} \frac{\rd \om_j}{2 \p (1 + \faq (\om_j))}
             \prod_{k=1}^{y_j - 1} (\om_j - \x_k + \i)
             \prod_{k = y_j + 1}^m (\om_j - \x_k) \biggr]
             \notag \\ &
        \frac{\det G(\om_j, \x_k)}
             {\prod_{1 \le j < k \le m}
                 (\x_k - \x_j) ( \om_j - \om_k - \i)} \epp
\end{align}
The rescaled auxiliary function $\fa$ satisfies the nonlinear integral
equation
\begin{equation} \label{nliexxx}
     \ln \fa (\la) = - 2\i \Phi + L \ln \biggl(
                       \frac{\la - \frac{\i}{2}}
		            {\la + \frac{\i}{2}} \biggr)
                     - \int_{\cal C} \frac{\rd \om}{\p}
		       \frac{\ln( 1 + \fa (\om))}{1 + (\la - \om)^2}
		       \epc
\end{equation}
and $\faq = 1/\fa$. The contour now surrounds the real axis
counterclockwise slightly below $\Im \la = \2$ and slightly above
$\Im \la = - \2$. The rescaled $G$ in the XXX limit is defined by an
integral equation on the same contour which reads
\begin{equation} \label{defgxxx}
     G(\la,\x) + \frac{1}{(\la - \x) (\la - \x - \i)} =
                   \int_{\cal C} \frac{d \om}{\p (1 + \fa (\om))} \,
                   \frac{G (\om,\x)}{1 + (\la - \om)^2} \epp
\end{equation}

For the following it is important to notice that (\ref{densintxxx})
and (\ref{defgxxx}) are exactly of the same form as in the finite-%
temperature case \cite{BGKS06}, where (for $m = 2, 3$) (\ref{defgxxx})
was used in order to factorize (\ref{densintxxx}). The argument did
not depend on the form of $\fa$ and therefore applies here in exactly
the same way. We just have to replace the finite-temperature auxiliary
function used in \cite{BGKS06} by the finite-size auxiliary function
(\ref{nliexxx}).

Let us review the results of \cite{BGKS06}. It was shown that the
inhomogeneous density matrix for $m$ up to 3 can be expressed in
terms of functions defined by single integrals. The most important
one is
\begin{equation} \label{defpsi}
     \psi (\x_1, \x_2) =
        \int_{\cal C} \frac{d \om \: G(\om, \x_1)}{\p (1 + \fa(\om))}
        \frac{1}{(\om - \x_2)(\om - \x_2 - \i)} \epp
\end{equation}
We suggest (see the conjecture below) that this is the only
transcendental function needed in the description of the inhomogeneous
density matrix for arbitrary $m$ and vanishing twist angle $\PH = 0$,
and that in this case the length dependence of the density matrix
enters through $\psi (\x_1, \x_2)$ alone. If we want to consider
non-zero $\PH$ we have to deal with another family of functions
\begin{equation}
        \phi_j (\x) = \int_{\cal C} \frac{d \om \: \om^{j-1} G(\om, \x)}
                                 {\p (1 + \fa(\om))} \epc \qd
                                 j \in {\mathbb N}
\end{equation}
which where called moments in \cite{BGKS06}.

For a compact notation of our final formulae it turns out to be
useful to introduce certain combinations of the moments with
rational functions. We first of all note that in the thermodynamic
limit for zero twist angle the moments turn into polynomials in $\x$
of order $j - 1$,
\begin{equation}
     \lim_{1/L \rightarrow 0} \lim_{\PH \rightarrow 0} \phi_j (\x)
        = \phi^{(0)}_j (\x)
        = (- \i \6_k)^{j-1} \, \frac{2 \re^{\i k \x}}{1 + \re^{k}}
          \Big|_{k=0} \epp
\end{equation}
The `normalized moments',
\begin{equation}
     \ph_j (\x) = \phi_j (\x) - \phi^{(0)}_j (\x) \epc
\end{equation}
then vanish for $1/L, \PH \rightarrow 0$. We use them to define the
symmetric combinations
\begin{equation}
     \D_n (\x_1, \dots, \x_n) =
        \frac{\det (\ph_j (\x_k))\bigr|_{j, k = 1, \dots, n}}
             {\prod_{1 \le j < k \le n} \x_{kj}} \epc
\end{equation}
where the shorthand notation $\x_{kj} = \x_k - \x_j$ was employed.
The first moment $\ph_1$ is exceptional among the $\ph_j$ in that it
becomes trivial even for finite length if only the twist angle vanishes,
\begin{equation}
     \lim_{\PH \rightarrow 0} \ph_1 (\x) = 0 \epp
\end{equation}
It follows that
\begin{equation} \label{deltavan}
     \lim_{\PH \rightarrow 0} \D_j (\x) = 0 \epc \qd
        \text{for all $j \in {\mathbb N}$.}
\end{equation}
Instead of $\ps (\x_1, \x_2)$ we shall use the closely related
expression
\begin{equation} \label{defgamma}
     \g (\x_1, \x_2) = \bigl[1 + (\x_1 - \x_2)^2\bigr]
                       \ps (\x_1, \x_2) - 1
\end{equation}
in terms of which our final formulae look neater. We also define
\begin{equation} \label{defgamma0}
     \g_0 (\x_1, \x_2) = \lim_{\PH \rightarrow 0} \g (\x_1, \x_2) \epp
\end{equation}

All density matrix elements for $m = 1, 2, 3$ can be written in terms
of these functions. A complete list can be found in the appendix of
\cite{BGKS06} where the functions $\g (\x_j, \x_k)$ and $\D_n (\x_1,
\dots, \x_n)$ have to be inserted according to our definitions above.
To give examples let us only recall the expressions for the
emptiness formation probabilities here,
\begin{subequations}
\begin{align} \label{p2and3}
     {D_L}^{11}_{11} (\x_1, \x_2) = & \frac{1}{4}
        - \frac{1}{12} \g (\x_1, \x_2)
        + \frac{1}{4} (\D_1 (\x_1) + \D_1 (\x_2))
        + \frac{1}{6} \D_2 (\x_1, \x_2)
	\epc \\[2ex]
     {D_L}^{111}_{111} (\x_1, \x_2, \x_3) = & \frac{1}{24}
        + \frac{1 - \x_{13} \x_{23}}{24 \x_{13} \x_{23}} \,
          \g (\x_1, \x_2) \notag \\[1ex] &
        + \frac{1 + 5 \x_{12} \x_{13}}{40 \x_{12} \x_{13}} \D_1 (\x_1)
        + \frac{1 + 2 \x_{13} \x_{23}}{24 \x_{13} \x_{23}}
          \D_2 (\x_1, \x_2) + \frac{1}{60} \D_3 (\x_1, \x_2, \x_3)
          \notag \\[1ex] &
        - \frac{3 + 2 \x_{12}^2 + 5 \x_{13} \x_{23}}
               {120 \x_{13} \x_{23}} \, \g (\x_1, \x_2) \D_1 (\x_3)
        + \text{cyclic perms.}
\end{align}
\end{subequations}
In the untwisted limit (\ref{deltavan}) applies and our result reduces
to
\begin{subequations}
\label{p23phi0}
\begin{align}
     {D_L}^{11}_{11} (\x_1, \x_2) = & \frac{1}{4}
        - \frac{1}{12} \g_0 (\x_1, \x_2) \epc \\[1ex]
     {D_L}^{111}_{111} (\x_1, \x_2, \x_3) = & \frac{1}{24}
        + \frac{1 - \x_{13} \x_{23}}
               {24 \x_{13} \x_{23}} \g_0 (\x_1, \x_2)
                    + \text{cyclic perms.}
\end{align}
\end{subequations}

Note that the only effect of taking the limit $1/L \rightarrow 0$ here 
is that the function $\g_0 (\x_1, \x_2)$ changes into its limiting form 
\begin{equation}
     \lim_{1/L \rightarrow 0} \g_0 (\x_1, \x_2)
        = 2 [1 + (\x_1 - \x_2)^2] {\cal K} (\x_1 - \x_2) - 1 \epc
\end{equation}
where
\begin{equation} \label{psizerozero}
     {\cal K} (x) = \i \: \6_x \ln \left[
                    \frac{\G \bigl( \tst{\2 + \frac{\i x}{2}} \bigr)
                          \G \bigl( \tst{1 - \frac{\i x}{2}} \bigr)}
                         {\G \bigl( \tst{\2 - \frac{\i x}{2}} \bigr)
			  \G \bigl( \tst{1 + \frac{\i x}{2}} \bigr)}
			  \right] \epp
\end{equation}
As in the temperature case a similar statement holds true for all
density matrix elements for $m = 1, 2$. The rational prefactors of
$\g_0 (\x_j, \x_k)$ are the same as in the thermodynamic limit.
For this reason it was conjectured in \cite{BGKS06} for the temperature
case that the exponential formula for general $m$ obtained in
\cite{BJMST05b} holds also for finite temperature. Further evidence
for this conjecture was supplied by the comparison of high-temperature
expansion data for the multiple integrals and for the conjectured
exponential formula for $m = 3, 4$. Regarding our results described
above it seems likely that the scope of the exponential formula is even
wider and that it also holds in the finite-length case under
consideration. Let us briefly recall how it looks like.

In order to obtain a convenient description of all density matrix
elements we shall resort to a notation that we borrowed from
\cite{BJMST06}\footnote{This definition was first introduced in
\cite{BJMST04a} and later modified in \cite{BJMST06}.}. We arrange
them into a column vector $h_m \in ({\mathbb C}^2)^{\otimes 2 m}$ with
coordinates labeled by $+, -$ instead of $1, 2$ according to the rule,
\begin{equation} \label{vecmat}
     h_m^{\e_1, \dots, \e_m, \bare_m, \dots, \bare_1}
         (\la_1, \dots, \la_m)
        = {D_L}^{(3 - \e_1)/2,  \dots, (3 - \e_m)/2}_%
            {(3 + \bare_1)/2, \dots, (3 + \bare_m)/2}
            (\xi_1, \dots, \xi_m) \cdot
            \prod_{j = 1}^m (- \bare_j) \epc
\end{equation}
where $\la_j = - \i \x_j$ for $j = 1, \dots, m$.
\begin{conjecture*}
The density matrix of a finite sub-chain of length $m$ of the
XXX chain of finite length $L$ is determined by the vector
\begin{align} \label{expform}
     h_m (\la_1, \dots, \la_m)
        & = \frac{1}{2^m} \re^{\Om_m (\la_1, \dots, \la_m)} \vs_m
            \epc \qqd \vs_m = \prod_{j=1}^m s_{j, \bar j} \epc
            \\[1ex] \label{defom}
     \Om_m (\la_1, \dots, \la_m) & = \frac{(-1)^{(m-1)}}{4}
        \int \int \frac{d \m_1}{2 \p \i} \frac{d \m_2}{2 \p \i}
        \frac{\g_0 (\i \m_1, \i \m_2) (\m_1 - \m_2)}
             {[1 - (\m_1 -\m_2)^2]^2} \\ & \notag \mspace{-72.mu} \times
        \tr_{\m_{1,2},2,2} \Bigl\{
           T \bigl( \tst{\frac{\m_1 + \m_2}{2}}; \la_1, \dots, \la_m
                      \bigr) \otimes \bigl[
           T (\m_1; \la_1, \dots, \la_m) \otimes
           T (\m_2; \la_1, \dots, \la_m) {\cal P}^- \bigr] \Bigr\} \epc
\end{align}
through (\ref{vecmat}). By the integral over $\m_1$, $\m_2$ it is
meant to take the residues at the poles $\la_1, \dots, \la_m$ of the
integrand.
\end{conjecture*}
For the notation we are referring to \cite{BJMST05b}\footnote{In fact,
the only difference between our formula (\ref{expform}), (\ref{defom})
and the result of \cite{BJMST05b} is in the function $\g_0$. In
\cite{BJMST05b} a function $\om$ was used which is related to $\g_0$ by
\[
     \omega(\la_1-\la_2) = \lim_{1/L \rightarrow 0}
                           \frac{\g_0 (i\la_1,i\la_2)}
			        {2\;(1-(\la_1-\la_2)^2)} \epp
\]
}: The vector
$s = \binom{1}{0} \otimes \binom{0}{1} - \binom{0}{1} \otimes
\binom{1}{0}$ is the spin singlet in ${\mathbb C}^2 \otimes
{\mathbb C}^2$. The vector spaces in $({\mathbb C}^2)^{\otimes 2m}$
are numbered in the order $1, 2, \dots, n, \bar n, \overline{n - 1},
\dots, \bar 1$. This defines $\vs_m$. $\2 {\cal P}^-$ is the projector
onto the one-dimensional subspace of ${\mathbb C}^2 \otimes
{\mathbb C}^2$ spanned by $s$.

In order to define the transfer matrices in the integrand in
(\ref{defom}) we first of all introduce an $L$-matrix $L (\la) \in
U (\mathfrak{sl}_2) \otimes \End {\mathbb C}^2$,
\begin{equation} \label{defl}
     L(\la) = \frac{\r (\la, d)}{2 \la + d}
             (2 \la + 1 + \Si^\a \otimes \s^\a) \epc
\end{equation}
where the $\Si^\a \in \mathfrak{sl}_2$ are a basis satisfying $[\Si^\a,
\Si^\be] = 2 \i \e^{\a \be \g} \Si^\g$, where $d$ is determined
by the Casimir element through $d^2 = (\Si^\a)^2 + 1$ and where
$\r (\la, d)$ satisfies the functional relation
\begin{equation}
     \r(\la, d) \r(\la - 1, d) = \frac{2 - 2 \la - d}{2 \la - d}
\end{equation}
(for more details see \cite{BJMST05b}). Then, for integer $z$, the
`transfer matrices'
\begin{multline}
     \tr_z T(\la; \la_1, \dots, \la_n) = \\
     \tr_z L_{\bar 1} (\la - \la_1 - 1) \dots
           L_{\bar n} (\la - \la_n - 1) L_n (\la - \la_n) \dots
           L_1 (\la - \la_1)
\end{multline}
entering (\ref{defom}) are defined by substituting the irreducible
representation of $U (\mathfrak{sl}_2)$ of dimension $z$ into the
definition (\ref{defl}) of the $L$-matrices. For non-integer
$z$ this can be analytically continued into the complex plane.
\section{A numerical case study}
Finally we would like to demonstrate that the formulae obtained
above are numerically efficient, at least for small $m$. The examples
we will be focusing on are the $zz$-correlation functions $\<\s_1^z
\s_2^z\>$ and $\<\s_1^z \s_3^z\>$ as functions of the chain length $L$
in the untwisted case $\PH = 0$. For these it is sufficient to know the
emptiness formation probability for $m = 2, 3$, since
\begin{equation}
     \<\s_1^z \s_2^z\> = 4 {D_L}^{11}_{11} - 1 \epc \qd
     \<\s_1^z \s_3^z\> = 8 {D_L}^{111}_{111} - 8 {D_L}^{11}_{11} + 1
                         \epp
\end{equation}
Hence, we have to perform the homogeneous limit $\x_j \rightarrow
\frac{\i}{2}$ in (\ref{p23phi0}), yielding
\begin{subequations}
\label{szszgamma}
\begin{align}
     \<\s_1^z \s_2^z\>
        = & - \tst{\frac{1}{3}}
	    \g_0 (\tst{\frac{\i}{2}}, \tst{\frac{\i}{2}})
        = \tst{\frac{1}{3}}
          - \tst{\frac{1}{3}}
	    \ps (\tst{\frac{\i}{2}}, \tst{\frac{\i}{2}}) \epc \qd
          \\[2ex]
     \<\s_1^z \s_3^z\>
        = & - \tst{\frac{1}{3}}
	    \g_0 (\tst{\frac{\i}{2}}, \tst{\frac{\i}{2}})
	  - \tst{\frac{1}{6}}
	    \g_{0,xx} (\tst{\frac{\i}{2}}, \tst{\frac{\i}{2}})
	  + \tst{\frac{1}{3}}
            \g_{0,xy} (\tst{\frac{\i}{2}}, \tst{\frac{\i}{2}})
	  \notag \\[1ex]
        = & \tst{\frac{1}{3}}
          - \tst{\frac{4}{3}}
	    \ps (\tst{\frac{\i}{2}}, \tst{\frac{\i}{2}})
	  - \tst{\frac{1}{6}}
	    \ps_{xx} (\tst{\frac{\i}{2}}, \tst{\frac{\i}{2}})
	  + \tst{\frac{1}{3}}
            \ps_{xy} (\tst{\frac{\i}{2}}, \tst{\frac{\i}{2}}) \epc
\end{align}
\end{subequations}
where we denote derivatives with respect to the first and second
argument by subscripts $x$ and $y$, respectively.

In order to calculate the functions $\ps$, $\ps_{xx}$ and
$\ps_{xy}$ on a computer we switch to the $\fb \fbq$-formulation
\cite{KluBatPea91,BoGo05} mentioned in section \ref{sec:bas}. For
real $x$ we define $\fb (x) = \fa (x + \frac{\i}{2})$ and $\fbq (x) =
\faq (x - \frac{\i}{2})$. Then \cite{KluBatPea91}
\begin{subequations}
\label{nliebbbar}
\begin{align}
     \ln \fb (x) = & L \ln ( \tgh (\p x/2))
        + \int_{- \infty}^\infty \frac{\rd y}{2 \p}
          {\cal K} (x - y) \ln \bigl(1 + \fb (y)\bigr)
	     \notag \\ & \mspace{162.mu}
        - \int_{- \infty}^\infty \frac{\rd y}{2 \p}
          {\cal K} (x - y + \i - \i 0) \ln \bigl(1 + \fbq (y)\bigr)
	     \displaybreak[0] \epc \\[1ex]
     \ln \fbq (x) = & L \ln ( \tgh (\p x/2))
        + \int_{- \infty}^\infty \frac{\rd y}{2 \p}
          {\cal K} (x - y) \ln \bigl(1 + \fbq (y)\bigr)
	     \notag \\ & \mspace{162.mu}
        - \int_{- \infty}^\infty \frac{\rd y}{2 \p}
          {\cal K} (x - y - \i + \i 0) \ln \bigl(1 + \fb (y)\bigr) \epc
\end{align}
\end{subequations}
where ${\cal K} (x)$ is defined by (\ref{psizerozero}). The function
$\ps$ and its derivatives can be expressed in terms of $\fb$ and
$\fbq$. For this purpose we also have to adapt the form of the function
$G(\la, \x)$. Following \cite{BoGo05} we define $g^{(\pm)}_\x (x)
= \pm G(x \pm \frac{\i}{2}, \x)$. These functions satisfy the linear
integral equations
\begin{subequations}
\label{defgsmall}
\begin{align}
     g^{(+)}_\x (x) = & \frac{- \p}{\ch(\p (\x - x))}
        + \int_{- \infty}^\infty \frac{\rd y \: g^{(+)}_\x (y)}
	  {2 \p (1 + \fb^{-1} (y))}
          {\cal K} (x - y) \notag \\ & \mspace{162.mu}
        - \int_{- \infty}^\infty \frac{\rd y \: g^{(-)}_\x (y)}
	  {2 \p (1 + \fbq^{-1} (y))}
          {\cal K} (x - y + \i - \i 0) \epc \\[1ex]
     g^{(-)}_\x (x) = & \frac{- \p}{\ch(\p (\x - x))}
        + \int_{- \infty}^\infty \frac{\rd y \: g^{(-)}_\x (y)}
	  {2 \p (1 + \fbq^{-1} (y))}
          {\cal K} (x - y) \notag \\ & \mspace{162.mu}
        - \int_{- \infty}^\infty
	  \frac{\rd y \: g^{(+)}_\x (y)}{2 \p (1 + \fb^{-1} (y))}
          {\cal K} (x - y - \i + \i 0) \epp
\end{align}
\end{subequations}
Using $\fb$, $\fbq$ and $g^{(\pm)}_\x$ we can express the function
$\ps (\x_1, \x_2)$ as
\begin{equation}
     \ps (\x_1, \x_2) = 2 {\cal K} (\x_1 - \x_2) 
        + \int_{- \infty}^\infty \frac{\rd x}{\ch(\p (\x_2 - x))}
	  \biggl[ \frac{g^{(+)}_{\x_1} (x)}{1 + \fb^{-1} (x)} +
                  \frac{g^{(-)}_{\x_1} (x)}{1 + \fbq^{-1} (x)} \biggr]
		  \epp
\end{equation}
This formulation is now convenient for the numerical evaluation
of the $zz$-correlation functions (\ref{szszgamma}) for which we need
$\ps (\frac{\i}{2}, \frac{\i}{2})$, $\ps_{xx} (\frac{\i}{2},
\frac{\i}{2})$ and $\ps_{xy} (\frac{\i}{2}, \frac{\i}{2})$. For the
expansion of the kernel function we can use
\begin{equation}
     {\cal K} (x)
        = 2 \sum_{k=0}^\infty (-1)^k \z_a (2k + 1) x^{2k} \, \epc
\end{equation}
where $\z_a (x) = \sum_{k=1}^\infty (-1)^{k+1} /k^x$ is the alternating
zeta series\footnote{For $\Re x > 1$ the Riemann zeta function $\z (x)$
and $\z_a (x)$ are related by $\z_a (x) = (1 - 2^{1 - x}) \zeta (x)$.}.
Then
\begin{align} \label{psihom}
     \ps (\tst{\frac{\i}{2}}, \tst{\frac{\i}{2}}) = & 4 \ln 2
        + \int_{- \infty}^\infty \frac{\i \rd x}{\sh(\p (x + \i 0))}
	  \biggl[ \frac{g^{(+)}_{\i/2} (x)}{1 + \fb^{-1} (x)} +
                  \frac{g^{(-)}_{\i/2} (x)}{1 + \fbq^{-1} (x)} \biggr]
		  \displaybreak[0] \epc \notag \\[1ex]
     \ps_{xy} (\tst{\frac{\i}{2}}, \tst{\frac{\i}{2}}) = & 6 \z (3)
        + \int_{- \infty}^\infty
	  \frac{\rd x \: \i \p \ch(\p x)}{\sh^2 (\p (x + \i 0))}
	  \biggl[ \frac{{g^{(+)}_{\i/2}}' (x)}{1 + \fb^{-1} (x)} +
                  \frac{{g^{(-)}_{\i/2}}' (x)}{1 + \fbq^{-1} (x)} \biggr]
		  \displaybreak[0] \epc \notag \\[1ex]
     \ps_{xx} (\tst{\frac{\i}{2}}, \tst{\frac{\i}{2}}) = & - 6 \z (3)
        + \int_{- \infty}^\infty \frac{\i \rd x}{\sh(\p (x + \i 0))}
	  \biggl[ \frac{{g^{(+)}_{\i/2}}'' (x)}{1 + \fb^{-1} (x)} +
                  \frac{{g^{(-)}_{\i/2}}'' (x)}{1 + \fbq^{-1} (x)}
		  \biggr] \epc
\end{align}
where the primes denote derivatives with respect to $\x$.
\begin{figure}
\includegraphics[height=100.mm,angle=0]{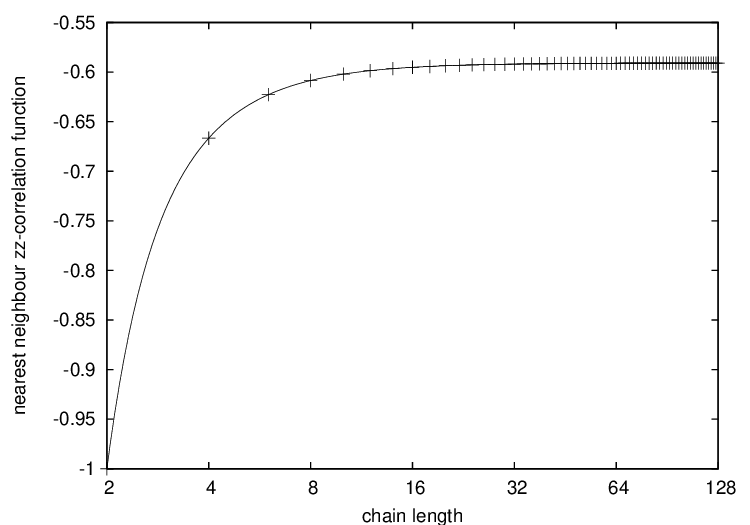}
\caption{\label{fig:fine2} Nearest neighbour $zz$-correlation function
for chains of even length $L$, solid curve represents the analytic
continuation to arbitrary real and positive $L$ as defined by
our integral representation.} 
\end{figure}
Using (\ref{defgsmall}) and (\ref{psihom}) in (\ref{szszgamma}) we
calculated the nearest and next-to-nearest neighbour $zz$-correlators
numerically. The non-linear integral equations~(\ref{nliebbbar}) as
well as the linear integral equations~(\ref{defgsmall}) were solved
iteratively in Fourier space. The derivatives of $g^{(\pm)}_\x$ with
respect to $\x$ were computed by solving the integral equations obtained
from (\ref{defgsmall}) by taking the derivatives with respect to $\x$.
\enlargethispage{-5ex}

\begin{figure}
\includegraphics[height=100.mm,angle=0]{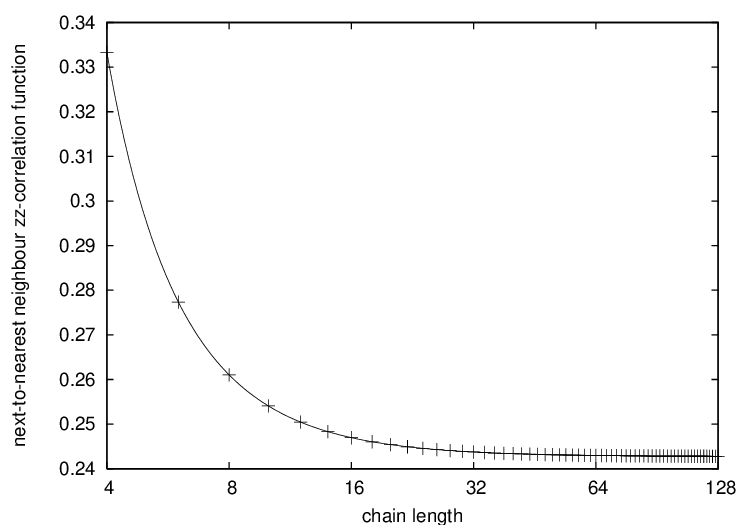}
\caption{\label{fig:fine3} Next-to-nearest neighbour $zz$-correlation
function for chains of even length $L$, solid curve represents the
analytic continuation to arbitrary real and positive $L$ as defined by
our integral representation.} 
\end{figure}

Figure \ref{fig:fine2} shows the nearest neighbour correlator and
figure~\ref{fig:fine3} the next-to-nearest neighbour correlator. Since
our model represents an antiferromagnet the former must be negative and
the latter positive. The weakening of the correlation with growing
length can be attributed to what is called `quantum frustration'
in condensed matter physics. Let us illustrate this notion with an
example. The (unnormalized) ground states of the XXX~chain of
length 2 and 4, respectively, are
\begin{align} \label{gs24}
     |gs\>_2 = & \underbrace{|\ab \auf\> - |\auf \ab\>}_{\text{N\'eel}}
               \epc \notag \\[1ex]
     |gs\>_4 = & \underbrace{2 |\ab \auf \ab \auf\> +
                 2 |\auf \ab \auf \ab\>}_{\text{N\'eel}}
             - \underbrace{|\ab \ab \auf \auf\> - |\auf \ab \ab \auf\>
	         - |\auf \auf \ab \ab\>
		 - |\ab \auf \auf \ab\>}_{\text{QM frustration}} \epp
\end{align}
The states with alternating up and down spins on consecutive sites
are called N\'eel states. They are a `classical caricature' of
an antiferromagnet. The $zz$-correlators in a N\'eel state are
$\<\s_1^z \s_{n+1}^z\> = (-1)^n$ and realize `perfect antiferromagnetic
order'. Such type of order is realized in the ground state of the
XXX chain only for $L = 2$, where the correlations look like in
the classical case (see figure~\ref{fig:fine2}) and where the
next-to-nearest neighbour correlator is not defined. For $L = 4$, as
can be seen from the ground state wave function (\ref{gs24}), there is
a certain probability to have parallel spins on neighbouring sites and
antiparallel spins on next-to-nearest neighbour sites. This reduces
the correlations in both cases. For growing chain length the `N\'eel
order' is even more frustrated, e.g.\ three or
four parallel spins appear, and the correlations are further reduced,
which explains the monotonous behaviour of our curves in figures~%
\ref{fig:fine2} and \ref{fig:fine3}.

It follows by inspection of equation~(\ref{nliebbbar}) that $\fb (x)$
and $\fbq (x)$ vanish in the thermodynamic limit $L \rightarrow
\infty$. Hence, the integrals in (\ref{psihom}) all vanish and
$\ps (\frac{\i}{2},\frac{\i}{2}) = 4 \ln 2$, $\ps_{xy} (\frac{\i}{2},
\frac{\i}{2}) = 6 \z (3)$, $\ps_{xx} (\frac{\i}{2},\frac{\i}{2}) =
- 6 \z (3)$. Inserting this into (\ref{szszgamma}) we obtain
\begin{align} \label{asymp23}
     \lim_{L \rightarrow \infty}
     \<\s_1^z \s_2^z\> = & \tst{\frac{1}{3}} - \tst{\frac{4}{3}} \ln 2
                         \notag \\[1ex]
     \lim_{L \rightarrow \infty}
     \<\s_1^z \s_3^z\> = & \tst{\frac{1}{3}} - \tst{\frac{16}{3}} \ln 2
                         + 3 \z (3)
\end{align}
The first equation is a corollary to Hulth\'en's classical result
\cite{Hul38} on the ground state energy per site of the XXX chain,
and the second one is a well known result due to Takahashi
\cite{Takahashi77} which was reproduced from the multiple integral
formula of Jimbo et al.\ \cite{JMMN92} by Boos and Korepin \cite{BoKo01}.
Here we have calculated the correlation functions for finite chain
length as correction to the asymptotic values (\ref{asymp23}).
The $zz$-correlators for $L = 4$, far away from these asymptotic
values, can be easily obtained from the ground state wavefunction
$|gs\>_4$ in (\ref{gs24}), $\<\s_1^z \s_2^z\> = - \frac{2}{3}$ and
$\<\s_1^z \s_3^z\> = \frac{1}{3}$. It is remarkable that these values
are reproduced from our integral equations to 13 digits precision
\begin{table}
  \centering
  \begin{tabular}{@{}rll@{}}
    \toprule
    $L$ & $\langle\sigma^z_1\sigma^z_2\rangle$
        & $\langle\sigma^z_1\sigma^z_3\rangle$\\
    \midrule
    2        & $-1.00000000000000$ & ---\\
    4        & $-0.66666666666667$ & $0.33333333333333$ \\
    8        & $-0.60851556815620$ & $0.26103720534839$ \\
    16       & $-0.59519136338473$ & $0.24696584167998$ \\
    32       & $-0.59193864328956$ & $0.24374937989865$ \\
    64       & $-0.59113127886152$ & $0.24297329183505$ \\
    128      & $-0.59092994011745$ & $0.24278223127753$ \\
    256      & $-0.59087965782193$ & $0.24273481483257$ \\
    512      & $-0.59086709385781$ & $0.24272300601642$ \\
    1024     & $-0.59086395383499$ & $0.24272006021644$ \\
    $\infty$ & $-0.59086290741326$ & $0.24271907982574$ \\
    \bottomrule
  \end{tabular}
\caption{\label{tab:finedata} $zz$-correlators as functions of the
system size.}
\end{table}%
without too much effort (see table \ref{tab:finedata}).

\section{Conclusions}
We have obtained a multiple integral formula (\ref{densint}) for
the zero temperature limit of the density matrix of a finite
segment of the XXZ chain which holds for every even chain length
$L$ and is again of the same form as the formerly known formula for
finite temperature but infinite length. The multiple integrals are
`parameterized' by a pair of functions $\fa (\la)$, $G(\la, \x)$
which fix their physical meaning. For one such pair we obtain the
finite-temperature density matrix for another pair the ground state
density matrix for the finite chain, the only difference being the
driving term in the non-linear integral equation~(\ref{nlie})
for $\fa (\la)$. In the thermodynamic limit for vanishing twist angle
(the zero temperature limit for vanishing magnetic field) the
integrands in the multiple integrals turn into explicit functions
and the formulae of Jimbo et al. \cite{JMMN92,JiMi96} are recovered.

Even the linear integral equations for $G(\la, \x)$ are of the same
form in the temperature case and in the finite-length case. Since
only this form was relevant for the reduction of the multiple
integrals to sums over products of single integrals (the `factorization')
in the finite-temperature case \cite{BGKS06}, a similar factorized form
of the short range correlations is valid for finite length and the
conjecture formulated for the general finite-temperature case in
\cite{BGKS06} is likely to hold in the finite-length case as well
(see (\ref{expform})). The phenomenon of factorization of correlation
functions, first observed in \cite{BoKo01} and rather well understood
in the thermodynamic limit for zero temperature and vanishing
external field by now \cite{BJMST05b}, may turn out to be valid in
a much broader context (compare the discussion in the summary of
\cite{BHMMOZ06}) and may even turn out to be a general characteristic
of quantum integrable models related to the Yang-Baxter equation.

In this work we have concentrated on the density matrix, since we
wanted to test if the factorization scheme of \cite{BGKS06} also
works in the finite-length case. We have seen that this is indeed
the case. Concerning the multiple integral representation we have no
doubt that similar formulae as derived for the two-point functions
\cite{GHS05} and for a generating function of the $zz$-correlation
functions \cite{GKS04a} for finite temperatures in the thermodynamic
limit also hold in the finite length case. It will be interesting
to compare the formulae for the generating function obtainable by
using the auxiliary function $\fa$ and our function $G$ with the
result of \cite{KMST05a}, where another multiple integral for the
finite-length system was derived.%
\\[1ex]{\bf Acknowledgement.}
The authors are indebted to H.~Boos, H.~Frahm, M.~Karbach, A.~Seel,
F.~Smirnov and J.~Suzuki for stimulating discussions. JD and NPH
acknowledge financial support by the DFG-funded research training
group 1052 -- `representation theory and its applications'.


\begin{thebibliography}{10}

\bibitem{Babook}
R.~J. Baxter, \emph{Exactly Solved Models in Statistical Mechanics}
(Academic Press, London, 1982).

\bibitem{BGKS06}
H.~Boos, F.~G\"ohmann, A.~Kl\"umper and J.~Suzuki,
\emph{Factorization of multiple integrals representing the density
matrix of a finite segment of the {H}eisenberg spin chain},
J. Stat. Mech.  (2006) P04001.

\bibitem{BJMST04a}
H.~Boos, M.~Jimbo, T.~Miwa, F.~Smirnov and Y.~Takeyama,
\emph{A recursion formula for the correlation functions of an
inhomogeneous {XXX} model}, Algebra and Analysis \textbf{17} (2005) 115.

\bibitem{BJMST06}
---, \emph{Algebraic representation of correlation functions in
integrable spin chains}, Ann. Henri Poincar\'e \textbf{7} (2006) 1395.

\bibitem{BJMST05b}
---, \emph{Density matrix of a finite sub-chain of the {H}eisenberg
anti-ferromagnet}, Lett. Math. Phys. \textbf{75} (2006) 201.

\bibitem{BJMST04b}
---, \emph{Reduced $q${KZ} equation and correlation functions of
the {XXZ} model}, Comm. Math. Phys. \textbf{261} (2006) 245.

\bibitem{BoKo01}
H.~E. Boos and V.~E. Korepin,
\emph{Quantum spin chains and {R}iemann zeta function with odd
arguments}, J. Phys. A \textbf{34} (2001) 5311.

\bibitem{BoGo05}
M.~Bortz and F.~G\"ohmann,
\emph{Exact thermodynamic limit of short range correlation functions
of the antiferromagnetic {XXZ} chain at finite temperatures},
Eur. Phys. J. B \textbf{46} (2005) 399.

\bibitem{BHMMOZ06}
S.~Boukraa, S.~Hassani, J.-M. Maillard, B.~M. McCoy, W.~P. Orrick and
N.~Zenine, \emph{Holonomy of the {I}sing model form factors},
J. Phys. A \textbf{40} (2007) 75.

\bibitem{GHS05}
F.~G\"ohmann, N.~P. Hasenclever and A.~Seel,
\emph{The finite temperature density matrix and two-point correlations
in the antiferromagnetic {XXZ} chain}, J. Stat. Mech.  (2005) P10015.

\bibitem{GKS04a}
F.~G\"ohmann, A.~Kl\"umper and A.~Seel,
\emph{Integral representations for correlation functions of
the {XXZ} chain at finite temperature},
J. Phys. A \textbf{37} (2004) 7625.

\bibitem{GKS05}
---, \emph{Integral representation of the density matrix of the
{XXZ} chain at finite temperature}, J. Phys. A \textbf{38} (2005) 1833.

\bibitem{GoKo00}
F.~G\"ohmann and V.~E. Korepin,
\emph{Solution of the quantum inverse problem},
J. Phys. A \textbf{33} (2000) 1199.

\bibitem{Hul38}
L.~Hulth\'en, \emph{{\"Uber das Austauschproblem eines Kristalles}},
Arkiv f\"or Matematik, Astronomi och Fysik \textbf{26A} (1938) 1.

\bibitem{JMMN92}
M.~Jimbo, K.~Miki, T.~Miwa and A.~Nakayashiki,
\emph{Correlation functions of the {XXZ} model for {$\Delta < - 1$}},
Phys. Lett. A \textbf{168} (1992) 256.

\bibitem{JiMi96}
M.~Jimbo and T.~Miwa, \emph{Quantum {KZ} equation with $|q| = 1$ and
correlation functions of the {XXZ} model in the gapless regime},
J. Phys. A \textbf{29} (1996) 2923.

\bibitem{KMST05a}
N.~Kitanine, J.~M. Maillet, N.~A. Slavnov and V.~Terras,
\emph{Master equation for spin-spin correlation functions of the
{XXZ} chain}, Nucl. Phys. B \textbf{712} (2005) 600.

\bibitem{KMT99a}
N.~Kitanine, J.~M. Maillet and V.~Terras, \emph{Form factors of
the {XXZ} {Heisenberg} spin-$\frac{1}{2}$ finite chain},
Nucl. Phys. B \textbf{554} (1999) 647.

\bibitem{KMT99b}
---, \emph{Correlation functions of the {XXZ} {H}eisenberg
spin-$\frac{1}{2}$ chain in a magnetic field},
Nucl. Phys. B \textbf{567} (2000) 554.

\bibitem{KluBat90}
A.~Kl\"umper and M.~T. Batchelor,
\emph{An analytic treatment of finite-size corrections of the spin-1
antiferromagnetic {XXZ} chain}, J. Phys. A \textbf{23} (1990) L189.

\bibitem{KluBatPea91}
A.~Kl\"umper, M.~T. Batchelor and P.~A. Pearce,
\emph{Central charges of the 6- and 19-vertex models with twisted
boundary conditions}, J. Phys. A \textbf{24} (1991) 3111.

\bibitem{KBIBo}
V.~E. Korepin, N.~M. Bogoliubov and A.~G. Izergin,
\emph{Quantum Inverse Scattering Method and Correlation Functions}
(Cambridge University Press, 1993).

\bibitem{Suzuki85}
M.~Suzuki, \emph{Transfer-matrix method and {Monte Carlo} simulation
in quantum spin systems}, Phys. Rev. B \textbf{31} (1985) 2957.

\bibitem{SuIn87}
M.~Suzuki and M.~Inoue, \emph{The {ST}-transformation approach to
analytic solutions of quantum systems. {I}.\ {G}eneral formulations
and basic limit theorems}, Prog. Theor. Phys. \textbf{78} (1987) 787.

\bibitem{Takahashi77}
M.~Takahashi,
\emph{Half-filled {Hubbard} model at low temperature},
J. Phys. C \textbf{10} (1977) 1289.

\end{thebibliography}

\end{document}